# End-to-End Speaker-Dependent Voice Activity Detection *

Yefei Chen, Shuai Wang, Yanmin Qian, Kai Yu

MoE Key Lab of Artificial Intelligence SpeechLab, Department of Computer Science and Engineering

Shanghai Jiao Tong University, 200000, China

**Abstract:** Voice activity detection (VAD) is an essential pre-processing step for tasks such as automatic speech recognition (ASR) and speaker recognition. A basic goal is to remove silent segments within an audio, while a more general VAD system could remove all the irrelevant segments such as noise and even unwanted speech from non-target speakers. We define the task, which only detects the speech from the target speaker, as speaker-dependent voice activity detection (SDVAD). This task is quite common in real applications and usually implemented by performing speaker verification (SV) on audio segments extracted from VAD. In this paper, we propose an end-to-end neural network based approach to address this problem, which explicitly takes the speaker identity into the modeling process. Moreover, inference can be performed in an online fashion, which leads to low system latency. Experiments are carried out on a conversational telephone dataset generated from the Switchboard corpus. Results show that our proposed online approach achieves significantly better performance than the usual VAD/SV system in terms of both frame accuracy and F-score. We also used our previously proposed segment-level metric for a more comprehensive analysis.

## 1. Introduction

Voice activity detection (VAD) [1], one of the most critical techniques of speech signal processing, is used to separate speech from non-speech segments within audio. VAD is usually applied as a pre-processing step for various speech processing tasks such as automatic speech recognition (ASR), speech synthesis, speaker recognition and voice over internet protocol (VoIP). The quality of VAD directly affects the performance of the subsequent tasks.

In traditional VAD systems, the non-speech parts are usually composed of silence and noises, while in this work we also incorporate the speech from unwanted speakers. This is quite common in real applications, for example, voice assistants may only need to reply to a particular speaker's commands, or in a conversational environment, speech from non-target speakers should be regarded as non-speech. The problem addressed is termed as speaker-dependent voice activity detection (SDVAD), which is an extension of the conventional VAD task. In this setting, we only want to detect the speech from a target speaker, so silence, noises, or speech from a non-target speaker will all be ignored. A naive approach to this task has two steps: (1) Generate speech segments using an ordinary VAD system (2) Perform speaker verification on the obtained speech segments to filter out the target speaker. However, this approach is performed in an offline manner and suffers from high latency.

Traditional VAD algorithms can be divided into two categories, feature-based methods and model-based methods. Regarding feature-based methods, different acoustic features are first extracted such as time domain energy [2], zero-crossing rate [3] and pitch [4], then simple detection scheme such as threshold comparison is applied. Regarding model-based methods, separate statistical models were trained to represent speech and non-speech segments by different probability distributions, where likelihood ratio between the two models is used as a decision threshold. Models such as Gaussian Mixture Model (GMM) [5] and Hidden Markov Model (HMM) [6] were investigated in the literature. Instead of using likelihood ratio based methods,

* This work has been supported by the National Key Research and Development Program of China under Grant No.2017YFB1302402 and the China NSFC projects (No. 61603252 and No. U1736202). Experiments have been carried out on the PI supercomputer at Shanghai Jiao Tong University. Yanmin Qian and Kai Yu are the corresponding authors.

directly training a binary classifier to discriminate speech and non-speech is more popular in current VAD systems. Classifiers such as Support Vector Machine (SVM) [7] and deep neural networks are trained to output the posteriors for each frame directly.

Recently, deep learning approaches have been successfully applied to many tasks including VAD. For VAD in complex environments, DNN has better modeling capabilities than traditional methods [8], recurrent neural network (RNN) and long short-term memory (LSTM) can better model long-term dependencies between inputs [9][10] and convolutional neural network (CNN) can generate better features for VAD training [11].

In order to tackle the speaker-dependent VAD problem, we propose a neural network based system which explicitly integrates the speaker identity information into the modeling process in this paper. On top of the normal spectral features such as filter banks (Fbank), speaker embeddings (*i*-vector) from the target speaker are also taken as input. If the current frame represented by the spectral feature is speech and comes from the speaker characterized by the *i*-vector, then the label will be 1, else it will be 0. Compared to the decoupled two-stage VAD / SV approach, our proposed model optimizes an end-to-end system directly against the final goal. Experiments are carried out on an artificial conversational dataset generated from the Switchboard database and results show that compared to the offline VAD / SV approach, our proposed online approach could achieve better performance with negligible latency since the prediction is generated at each frame.

The rest of this paper is organized as follows. In section 2, we introduce neural network based VAD. Section 3 describes the details of our proposed end-to-end speaker-dependent VAD architecture. In section 4, experimental results and analysis are provided. Discussion and conclusion are given in section 5.

## 2. Neural network based VAD

### 2.1 DNN-based VAD system

As shown in [8], the DNN based VAD systems not only outperform the traditional model-based systems but also have a low detection complexity [12]. A typical DNN based VAD system trains a frame-based binary classifier to classify each frame into two classes: speech and non-speech. Conventionally, the frame-wise input for the DNN is concatenated with its context as $\mathbf{O}_t$:

$$\mathbf{O}_t = [\mathbf{x}_{t-r}, \ldots, \mathbf{x}_{t-1}, \mathbf{x}_t, \mathbf{x}_{t+1}, \ldots, \mathbf{x}_{t+r}] \quad (1)$$

Where $\mathbf{x}_t$ is *t*-th frame and $r$ is the length of context extension. DNN is optimized by the cross entropy criterion. For each frame, classification is performed by a comparison among posterior probabilities of the two classes.

### 2.2 LSTM-based VAD system

LSTM is capable at modeling sequences and capturing long-range dependencies in a sequence of features. In its core it is comprised of special units called memory blocks. Each memory block contains an *input gate*, an *output gate* and a *forget gate*, which enables the model to memorize information for a short or long duration. The LSTM structure can effectively use a context to model the input acoustic features sequentially.

The LSTM network computes a mapping from an input sequence $\mathbf{x} = [\mathbf{x}_1, \mathbf{x}_2, \ldots, \mathbf{x}_T]$ to an output sequence $\mathbf{y} = [\mathbf{y}_1, \mathbf{y}_2, \ldots, \mathbf{y}_T]$. More details of this architecture could be referred from [13].

If applied to VAD, a LSTM based system outputs predictions frame by frame, but each prediction of the current frame partially depends on its history. The training criterion is the same as for DNN.

## 3. Speaker-dependent VAD

### 3.1 Related work

For speaker-dependent VAD, some previous studies [14][15] used microphone array to track target speaker. Authors in [16] also take the speaker identity into consideration for VAD, the VAD system they used was based on a Gaussian mixture model, while an additional GMM adapted to the target speaker is used to represent the speaker identity. However, it should be noted that we have a different experimental setting and we aim to solve different problems. In their research, speech from other speakers appeared as background noise, while in our setting we consider a conversational scenario, where speech from different speakers do not overlap. Another scenario is the use of smart speakers at home, where the speech recognition system will be disturbed by family member conversations. In general, for systems which only want to accept speech signal from a specific speaker, such a speaker-dependent voice activity detector is needed.

### 3.2 I-vector for speaker modeling

Speaker modeling plays a vital role in speech processing tasks such as speaker recognition, speaker diarization, speaker adaptation for speech recognition. In recent years, the factor analysis based *i*-vector systems achieved remarkable performance improvement for the speaker recognition task [17], such speaker representations are applied to other related tasks such as voice conversion [18] and speaker adaptive training for speech recognition [19].

Basically, *i*-vector is a low dimensional fixed-length representation of a speech utterance that preserves the speaker-specific information. For *i*-vector framework [20], the speaker and session dependent super-vector **M** (derived from UBM) is modeled as

$$\mathbf{M} = \mathbf{m} + \mathbf{Tw} \quad (2)$$

where **m** is a speaker and session independent super-vector, **T** is a low rank matrix that captures speaker and session variability, *i*-vector is the posterior mean of **w**.

### 3.3 Baseline system

As mentioned in the introduction, for the task of speaker-dependent VAD, an intuitive method will be a two-stage approach. First, ordinary VAD is used to detect all the speech segments without differentiating speakers, then we use a speaker verification system to pick out speech segments belonging to the target speaker. Thus, the baseline system is a combination of VAD and text-independent speaker verification, which will be termed as the VAD / SV approach in the rest of the paper.

In this work, DNN and LSTM based systems are trained for the VAD stage, while for the speaker verification part, we use a state-of-the-art text-independent speaker verification approach based on *i*-vector and Probabilistic Linear Discriminant Analysis (*i*-vector/PLDA framework).

### 3.4 End-to-end speaker-dependent VAD system (SDVAD)

According to the baseline system, the speaker verification stage is after obtaining VAD prediction results of the whole utterance, which increase system latency. Moreover, it does not directly optimize the ultimate goal of this task. Thus, we propose to introduce the speaker modeling jointly with the original VAD network to enable the model to give frame-level speaker-dependent predictions. Since the model is now trained in an end-to-end manner, the information of the data could be fully exploited to get a better system.

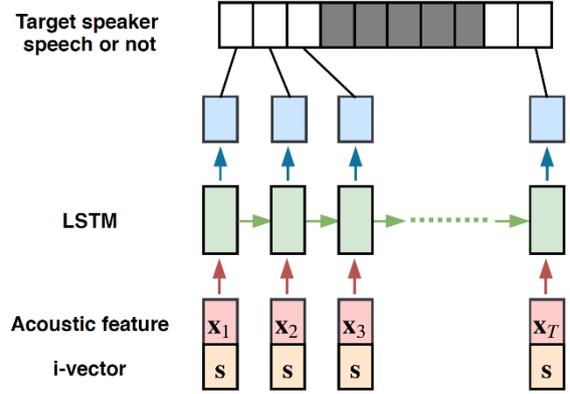

Figure 1: *LSTM based speaker-dependent VAD, the acoustic feature of each frame from one conversation is concatenated with i-vector from the target speaker.*

The proposed system is depicted in Figure 1, with the pretrained *i*-vector extractor, the *i*-vector of target speaker will be extracted from his specific speech. Then we concatenate the frame-level acoustic features and *i*-vector as a new input for the neural network. This is feasible both for the training and inference stages. For the training stage, the conversational data is well-annotated, thus speaker-specific data can be used in order to extract the corresponding *i*-vector. In the inference stage, it is reasonable to ask the users to first enroll their voice when they first use the system.

During the training process, only the speech part of the target speaker is regarded as a positive sample, while the speech part of non-target speakers and non-speech parts are regarded as negative examples. Therefore, the model is capable of directly outputting the final speaker-dependent predictions of each frame, without an extra speaker verification phase. The proposed speaker-dependent VAD system is an online system with negligible latency.

### 3.5 Post-processing and Feature Binning

VAD is different from common binary classification problems since the audio signal is characterized by continuity which means adjacent frames are highly correlated. The raw output of the model often contains many false transitions resulting in a "fragmentation problem" due to impulse noise and other interference. For frame-based classifier like DNN, this kind of problem is more obvious. So it is important to apply post-processing methods to smooth the raw output of models and reduce frequent and false transitions between speech and non-speech. Specifically for rule-based post-processing methods, a sliding window is used to reduce short term variations, then short duration segments are merged.

Most post-processing methods will add extra latency to online VAD system. In this paper, another method called feature binning is used to help solve the "fragmentation problem" in speaker-dependent VAD. The difference is that we try to smooth the input feature instead of the output of models. Regarding VAD, feature binning is accomplished by grouping the values into a fixed number of bins. The continuous value then gets replaced by a single value. As shown in Figure 2, we merge the input features of n frames adjacent without overlap using mean reduction. This procedure reduces the original frame size to a factor of $1/n$. Then each output prediction of model is repeated n times to correspond to the original feature of each frame. The latency caused by this method is negligible.

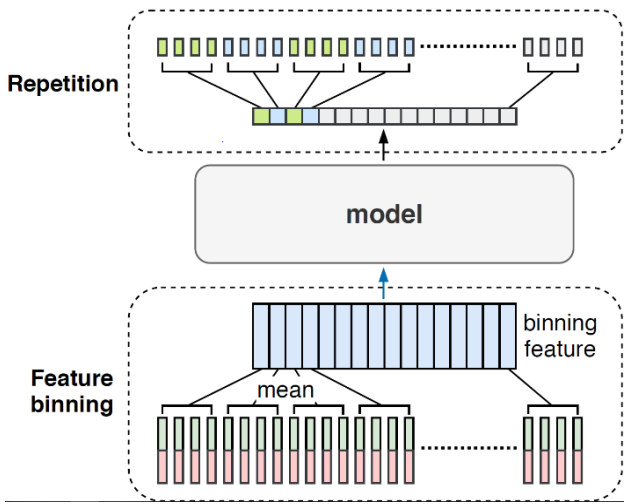

Figure 2: *Feature Binning for speaker-dependent VAD*

For the DNN model, the normal frame-extension is used to add context information and reduce false transitions in prediction results. For the LSTM model, we use feature binning to keep the continuity of speech and reduce the computation cost.

## 4. Experiments

### 4.1 Dateset

We conducted experiments on a conversational dataset generated from Switchboard corpus. After sifting out duplicate utterances and speakers with insufficient data, 250h of audio data of 500 speakers left, where each utterance contains only one speaker. Then we divided these filtered data into train, dev and test set. There are 450 speakers in the train set, 10 speakers in the development set and the remaining 40 speakers in the test set.

The generation process for the training data is as follows, (1) Prepare *i*-vectors for the speakers in the training set. (2) Randomly select the *i*-th utterance from the *s*-th speaker, termed as $utt_i^s$ and the *j*-th utterance from the *t*-th speaker, termed as $utt_j^t$, where $s \neq t$ Concatenate the two utterances $utt_i^s$ and $utt_j^t$ to get a new utterance $utt_{new}$, while treating speaker *s* as the target speaker. Only the speech of target speaker *s* is regarded as positive samples, while the speech of speaker *t* and non-speech are labeled as negative samples. (3) Concatenate the *i*-vector for speaker *s* to each frame of spectral features of $utt_{new}$ to formulate the final input for the neural network. The generation of the development data and test data is similar, while we assume the *i*-vector for the target speaker is obtained by an extra enrollment stage.

### 4.2 Features

For the *i*-vector extractor, 20-dimensional MFCCs with a frame-length of 25ms are extracted as front-end features. The UBM consist of 2048 component full covariance GMM and the dimension of extracted *i*-vector is 200. PLDA serves as a scoring back-end and compensates the channel distortion. The basic feature for all neural networks was 36-dimensional log mel filterbank extracted from 25ms frames with 10ms frame shifts. For the DNN model, the input layer was formed from a context window of 11 frames. The DNN consists of 2 hidden layers. LSTM model contains 2 hidden layers.

### 4. 3 Frame level Evaluation

Results for the frame-level evaluation are reported in terms of accuracy (ACC) and F-score (F1, harmonic mean of precision and recall), which are listed in Table 1.

Table 1: *ACC(%) and F-score(%) of different systems. VAD / SV means VAD followed by speaker verification, the two-stage baseline system while SDVAD indicates our proposed end-to-end speaker-dependent VAD system. "+ post" and "+ binning" represent applying post-processing and feature binning respectively, For post-processing, the size of sliding window is 10 frames. The size of feature binning is 4.*

| Systems | ACC(%) | F-score(%) |
|---|---|---|
| DNN VAD/SV | 81.71 | 81.01 |
| LSTM VAD/SV | **86.62** | 85.29 |
| LSTM VAD/SV+binning | 86.50 | **85.30** |
| DNN SDVAD | 83.41 | 77.68 |
| LSTM SDVAD | 88.31 | 85.71 |
| DNN SDVAD+post | 85.70 | 80.92 |
| LSTM SDVAD+post | 89.11 | 86.86 |
| LSTM SDVAD+binning | **94.42** | **93.22** |
| LSTM SDVAD+binning+post | 94.62 | 93.47 |

If without any pre or post-processing, it can be found that LSTM has better performance than DNN in both VAD / SV baseline and SDVAD system, which is attributable to its sequence modeling capability. For the LSTM SDVAD system, ACC and F-score of SDVAD system are slightly higher than VAD / SV baseline system, which means our proposed speaker-dependent VAD method is effective. The speaker-aware training paradigm of the proposed system leverages the speaker information to a good extent.

In order to solve the "fragmentation problem" and further improve the system's performance, rule-based post-processing and feature binning mentioned in Section 3.5 are applied to these systems. From the results one can see that post-processing can slightly improve the performance of DNN and LSTM SDVAD systems.

On the other hand, the proposed feature binning method can greatly benefit the LSTM-based SDVAD system, improving the ACC from 88.31% to 94.42% and can be further enhanced to 94.62% via post-processing. F-score has the same improvement as ACC.

Here we need to note, as the first stage of the baseline system, ordinary VAD can get good accuracy for speech/non-speech classification (no speaker distinction) without too many fragments. Feature binning does not have much impact on the first stage, so it can not improve the whole VAD / SV system. For the same reason, post-processing method can not bring improvement to VAD / SV systems so the results of VAD / SV with post-processing were not added to Table 1. The reason for the difference in performance between two process methods is that the post-processing operation does not affect the training process of the SDVAD, while the feature binning, as a pre-processing step, could be regarded as a part of the neural network, which helps the network fully exploit the information.

### 4.4 Segment level Evaluation

ACC and F-score are only indications of frame classification ability. We want to further investigate the performance of VAD / SV baseline and SDVAD systems at the segment-level. The evaluation metric $\mathcal{J}_{VAD}$ proposed in our previous work [21] is used here.

$\mathcal{J}_{VAD}$ contains four different sub-criteria, namely start boundary accuracy (SBA), end boundary accuracy (EBA), border precision (BP) and frame accuracy (ACC). ACC is the basic percentage of correctly recognized frames. SBA and EBA are indications of boundary-level accuracy. BP is a measure for the integrity of the VAD output segments. The harmonic mean of above four sub-criteria is defined as the segment level $\mathcal{J}_{VAD}$. The analysis is conducted from these four aspects. The detail $\mathcal{J}_{VAD}$ results are shown in Table 2.

Table 2: *segment level evaluation of different systems, $\mathcal{J}_{VAD}$(%) and 3 sub-criteria(%) of different systems are listed except ACC which has been shown in Table 1*

| Systems | SBA | EBA | BP | $\mathcal{J}_{VAD}$ |
|---|---|---|---|---|
| LSTM VAD/SV | 74.42 | 76.56 | 70.81 | **76.68** |
| LSTM SDVAD | 64.08 | 63.79 | 33.89 | 55.47 |
| LSTM SDVAD+binning | 71.73 | 73.67 | 61.74 | **73.66** |

For more intuitive comparison, only the LSTM model is used here. Compared with the VAD / SV baseline system, we can find that the original SDVAD system is limited by the "fragmentation problem". The prediction of SDVAD system may contain some false state transitions and fragments without any pre or post-processing. These fragments lead to an increase in the number of detected speech segments. Therefore, the BP evaluation is poor. Feature binning can reduce these false transitions efficiently. All segment level evaluation indicators have been improved and are close to the performance of the baseline system at segment level.

To better compare different systems, the prediction results on a test case are depicted in Figure 3. It can be observed that there are some fragments in the prediction results of SDVAD system and feature binning can effectively address the "fragmentation problem" in speaker-dependent VAD.

The VAD / SV system give some false alarms for non-target speakers, which is reasonable since VAD and SV are two decoupled stages, which could not be optimized towards the ultimate goal of the task.

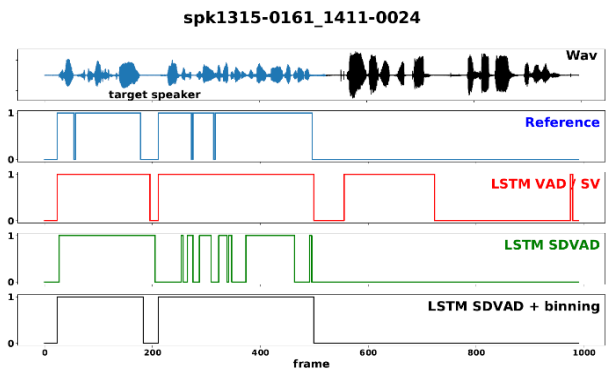

Figure 3: *Prediction of different systems*

## 5. Conclusion

In this paper, an end-to-end neural network based system is designed to address the speaker-dependent VAD problem, which aims to only detect the speech from a target speaker. Compared to the two-stage VAD / SV approach which suffers from high latency, our proposed end-to-end approach (SDVAD) directly takes the speaker information into the modeling process and could perform online predictions directly. Results of a series of experiments are reported in terms of common frame-level evaluation metrics and our previously proposed segment-level metric. For the frame-level evaluation, our proposed LSTM SDVAD system achieved significant performance improvement than the conventional VAD / SV system, from 86.62% to 94.42% in terms of frame accuracy. To address the "fragmentation problem", we introduce feature binning in the LSTM SDVAD system, which improves segment-level results significantly.